\documentclass[sigplan,10pt,screen]{acmart}

\copyrightyear{2026}
\acmYear{2026}
\setcopyright{cc}
\setcctype{by}
\acmConference[SOSP '26]{ACM SIGOPS 32nd Symposium on Operating Systems Principles}{September 29-October 02, 2026}{Prague, Czech Republic}
\acmBooktitle{ACM SIGOPS 32nd Symposium on Operating Systems Principles (SOSP '26), September 29-October 02, 2026, Prague, Czech Republic}
\acmDOI{10.1145/3830418.3843900}
\acmISBN{979-8-4007-2585-2/2026/09}

\usepackage{tikz}
\usepackage{amsmath}
\usepackage{booktabs}

\usepackage{amsthm}

\usepackage{xcolor}
\usepackage{enumitem}
\usepackage{graphicx}
\usepackage{subcaption}
\usepackage{hyperref}
\usepackage{mathtools,amssymb,latexsym,amsfonts,stmaryrd}
\usepackage{amsmath}
\usepackage{listings}
\usepackage{multirow}
\usepackage{pifont}
\usepackage[many]{tcolorbox} %
\usepackage{listings}
\usepackage{xspace}
\usepackage{algorithm}
\usepackage[noend]{algpseudocode}
\usepackage{mathpartir}
\usepackage[inference]{semantic}
\usepackage{wrapfig}

\definecolor{commentgreen}{RGB}{63,127,95}
\definecolor{beacon@phase}{RGB}{134, 153, 100}
\definecolor{beacon@body}{RGB}{80, 100, 80}
\definecolor{beacon@highlight}{RGB}{180, 70, 60}
\definecolor{beacon@bg}{RGB}{248, 248, 245}

\newcommand{\tool}{\textsc{StateLens}}
\newcommand{\parh}[1]{\noindent\textbf{#1}}
\newcommand{\parhs}[1]{\noindent\underline{\textit{#1}}}

\lstdefinelanguage{js}{
  basicstyle=\ttfamily\footnotesize,
  numberstyle=\footnotesize,
  numbers=left,
  numbersep=5pt,
  xleftmargin=7pt,
  tabsize=2,
  breaklines=true,
  breakatwhitespace=false,
  columns=fullflexible,
  keepspaces=true,
  showtabs=false,
  keywords={if, else, for, while, do, break, continue, return, class, const, static, let, function},
  sensitive=true,
  comment=[l]{//},
  escapeinside={(*@}{@*)},          %
  morecomment=[s]{/*}{*/},
  commentstyle=\color{commentgreen}\ttfamily,
  morestring=[b]',
  morestring=[b]",
  escapeinside={(*@}{@*)} %
}

\usepackage[capitalise]{cleveref}

\usepackage{parskip}

\lstdefinelanguage{beacon}{
  keywords=[1]{Phase},
  keywords=[2]{},
  sensitive=true,
  morecomment=[s]{**}{**},
}

\lstdefinestyle{beaconstyle}{
  language=beacon,
  basicstyle=\ttfamily\footnotesize\color{beacon@body},
  keywordstyle=[1]\ttfamily\color{beacon@phase},
  keywordstyle=[2]\ttfamily\color{beacon@highlight},
  commentstyle=\ttfamily\bfseries\color{beacon@body},
  moredelim=[s][\ttfamily\color{beacon@highlight}]{<}{>},
  backgroundcolor=\color{beacon@bg},
  frame=none,
  breaklines=true,
  breakatwhitespace=false,
  columns=fullflexible,
  keepspaces=true,
  showstringspaces=false,
}

\begin{document}

\title{State-Aware Fuzzing of JavaScript Engines with LLM-Guided Instrumentation}

\author{Wai Kin Wong}
\authornote{Equal contribution.}
\affiliation{%
  \institution{Hong Kong University of Science and Technology}
  \city{Hong Kong}
  \country{China}}
\email{wkwongal@cse.ust.hk}
\orcid{0000-0002-2583-0698}

\author{Dongwei Xiao}
\authornotemark[1]
\affiliation{%
  \institution{Hong Kong University of Science and Technology}
  \city{Hong Kong}
  \country{China}}
\email{dxiaoad@cse.ust.hk}
\orcid{0000-0002-4680-5715}

\author{Cheuk Tung Lai}
\affiliation{%
  \institution{VX Research Limited}
  \city{London}
  \country{United Kingdom}}
\orcid{0009-0002-0941-9791}
\email{darkfloyd@vxrl.hk}

\author{Ping Fan Ke}
\affiliation{%
  \institution{Singapore Management University}
  \city{Singapore}
  \country{Singapore}}
\orcid{0000-0002-4205-7801}
\email{pfke@smu.edu.sg}

\author{Shuai Wang}
\authornote{Corresponding author.}
\affiliation{%
  \institution{Hong Kong University of Science and Technology}
  \city{Hong Kong}
  \country{China}
}
\orcid{0000-0002-0866-0308}
\email{shuaiw@cse.ust.hk}

\begin{abstract}
The security of the modern web depends on the correctness of JavaScript (JS) engines, yet these complex systems remain vulnerable to high-impact bugs. A critical limitation of state-of-the-art fuzzers is the coverage plateau: once a fuzzer saturates the control-flow graph, edge coverage loses its ability to guide discovery. Because complex engine behaviors, such as JIT optimization tiers and hidden class transitions, often share identical edge coverage, standard coverage metrics are blind to the distinct internal states required to trigger deep errors.
To bridge this gap, we present \tool , a framework that employs Large Language Models (LLM) to automate the discovery of deep internal states. Blindly placing instrumentation probes at all states is infeasible due to the vast state space and the high runtime overhead. \tool\ introduces a novel agent-based reasoning pipeline that emulates the intuition of a security researcher. By iteratively traversing code and developer comments, our agents intelligently select high-value instrumentation targets, effectively separating logic-driving states from irrelevant data. This results in synthesizable, high-signal feedback probes that map the engine's hidden configurations. This instrumentation feeds a dual-feedback mechanism, effectively guiding the fuzzer toward unexplored engine semantics. Our evaluation confirms that \tool\ significantly outperforms state-of-the-art fuzzers and uncovers 68 new bugs.
\end{abstract}



\maketitle
\pagestyle{empty} %
\thispagestyle{empty} %

\section{Introduction}
\label{sec:introduction}

The security of today's software stack increasingly hinges on the security of JavaScript (JS) engines. Once confined to browsers, JS engines now underpin platforms ranging from web clients and server-side runtimes such as Node.js~\cite{nodejs}, Deno~\cite{deno}, and Bun~\cite{bunjsc} to serverless and edge platforms like AWS Lambda~\cite{awslambda} and Cloudflare Workers~\cite{cloudflaresandbox}. Their ubiquity makes them a prime attack surface: a single vulnerability can put billions of users or entire application stacks at risk through malicious web content alone. Indeed, nearly one-third of in-the-wild exploits in 2024 targeted JS engines~\cite{itw0day}.

Discovering deep vulnerabilities in JS engines remains challenging. \textit{Structural coverage} like line and branch coverage has long been the primary feedback signal for JS engine fuzzers~\cite{gross2023fuzzilli,han2019codealchemist}. However, reaching a code path is often insufficient to trigger a bug. Many high-severity JS-engine bugs stem from logic errors in internal state management and manifest only under specific runtime conditions. Even with identical code paths, semantically distinct internal states can lead to different outcomes.

Unlike structural coverage, which captures \emph{which code} executes, \emph{state coverage}~\cite{Li2024,aschermann2020ijon} distinguishes \emph{under what runtime conditions} that code executes. Prior approaches, however, are difficult to scale to JavaScript engines. They either require manual state selection, or are agnostic to the semantics of JS engines. These approaches fail to capture the complex, cross-component interactions that often harbor deep engine bugs.

The key challenge is deciding \emph{which} states to instrument in a vast internal state space. Exhaustive instrumentation is infeasible, yet sparse instrumentation leaves the fuzzer blind. We observe that developers have already encoded this knowledge in comments, design documents, historical bug reports, and debug assertions that describe which states are fragile and why. Given LLMs' ability for semantic understanding, they are well-suited to extract this knowledge and identify semantically meaningful states for instrumentation. We thus introduce \emph{\tool}, a framework that for the first time, enables state coverage for JS engines with the power of LLMs for automated state identification and instrumentation.

\emph{\tool} operates in three stages. In the \emph{analysis stage}, an LLM agent guided by a queryable knowledge base of developer artifacts performs iterative call-graph and data-flow traversal to identify semantically meaningful state expressions, transitions, and cross-component interactions. In the \emph{instrumentation stage}, it synthesizes lightweight, read-only probes that project these states into a compact shared-memory bitmap. During the \emph{fuzzing stage}, a dual-feedback mechanism integrates state coverage into the fuzzing loop alongside traditional edge coverage, retaining inputs that exercise previously unseen state combinations even when edge coverage reports no change. In summary, this paper makes the following contributions:
\begin{itemize}[leftmargin=*,topsep=3pt,itemsep=3pt]
  \item Conceptually, this work is the first to automatically identify and instrument JavaScript engine states as fuzzing feedback. For complex systems like JavaScript engines, reaching code/branch alone is insufficient to explore the rich semantic space in the engine. Furthermore, rather than relying on human experts to identify critical states for instrumentation, we propose to automatically mine such states with LLM agents to enable scalable state-aware fuzzing.
  \item Technically, we design and implement  \tool, a three-stage instrumentation pipeline that automatically identifies behavior-relevant state expressions from developer-authored artifacts and synthesizes validated probes in production JS engines, together with a dual-feedback adapter that integrates semantic state projections into the fuzzing loop alongside traditional edge coverage.
  \item Experimentally, we evaluate \tool\ on six major JS engines (V8, SpiderMonkey, JavaScriptCore, QuickJS, Hermes, and Escargot) and discover 68 new bugs in a long-running fuzzing campaign. In a 72-hour comparison with baseline fuzzers, \tool\ found 70\% more bugs than the best baseline fuzzer.
\end{itemize}

\section{Background}
\label{sec:background}

\parh{JS Engines as Critical System Infrastructure.}~JS engines have evolved from browser components into foundational infrastructure for modern computing systems. In desktop software, frameworks such as Electron embed engines like V8 to power widely deployed cross-platform applications, including Visual Studio Code, Slack, and Discord. In server environments, runtimes such as Node.js and Deno underpin backend services, cloud functions, and microservices. Major cloud providers further rely on managed JS runtimes to execute cloud and edge workloads, including AWS Lambda, Google Cloud Functions, and Cloudflare Workers. Vulnerabilities in JS engines have been repeatedly weaponized to achieve remote code execution (RCE) across fleets of Electron-based desktop applications ~\cite{ali2024rise, yang2025coindef}, while sandbox-escape bugs in server-side JS runtimes have threatened cross-tenant isolation in serverless cloud architectures ~\cite{alhamdan2023sanddriller}. The security and reliability of modern desktop and cloud systems therefore hinge on the robustness of JS engines, making effective fuzzing of these complex systems a critical priority.

\parh{JS Engines Architecture.}~The internal architecture of modern JS engines is inherently complex. These engines are highly stateful, multi-stage software systems comprising interpreters, multi-tier Just-In-Time (JIT) compilers, concurrent garbage collectors (GC), and WebAssembly runtimes. Moreover, their execution behavior can change drastically with the internal state. For instance, JS engines' memory management subsystems often employ sophisticated heap allocation and garbage collection strategies that are sensitive to current memory pressure and object models (e.g., hidden classes or shapes). This intricate interplay between dynamic compilation, memory management, and execution creates a vast space where vulnerabilities may arise only under highly specific, history-dependent sequences of runtime events.

\parh{Fuzzing JS Engines.}~Fuzzing has become a primary technique for vulnerability discovery in complex software ~\cite{jiang2025fuzzing, gong2025snowplow, mao2025nass}, and has been widely adopted for testing production JS engines~\cite{wang2023fuzzjita, dumpling}. However, directly applying general-purpose fuzzers like AFL ~\cite{aflcmin} is largely ineffective; such tools mutate raw bytes and rely on basic structural control-flow edges as mutation guidance, which is insufficient to navigate JS's massive syntactic input space or trigger deeply nested compilation states. Consequently, the state of the art has shifted toward domain-specific fuzzers, such as Fuzzilli ~\cite{gross2023fuzzilli}, DIE ~\cite{park2020fuzzing}, and Superion ~\cite{wang2019superion}, that operate on custom intermediate representations (e.g., FuzzIL) or carefully preserve semantic properties during mutation to generate valid programs. Yet, even with these specialized generators, uncovering deep bugs remains computationally expensive and many vulnerable states are still systematically missed.

\section{Motivation}

\label{sec:motivation}

Existing JS engine fuzzers commonly rely on structural coverage metrics, such as edge coverage, to guide input generation. In this section, we first use a real-world vulnerability to illustrate why structural coverage is insufficient. We then explain why existing approaches to state coverage do not scale to JS engines, before presenting observations that inform \tool's state coverage design.

\subsection{The Limits of Structural Coverage}
\label{subsec:limits_structural}

\parh{Structural Coverage Plateau.} Modern JS engines, such as V8 and SpiderMonkey, comprise millions of lines of C++ code. The same control-flow paths are repeatedly exercised under different object layouts, optimization tiers, heap conditions, and cache configurations.
Existing JS engine fuzzers~\cite{park2020fuzzing,gross2023fuzzilli,wang2024optfuzz,hlpfuzz} guide input generation with structural coverage metrics, such as edge coverage, retaining inputs that explore new control-flow edges. However, once these edges saturate, executions that reach new internal conditions may provide no new feedback.

\begin{lstlisting}[language=js, numbers=left]
function main(i) {
  class C { m() { return super.x; } }
  let rect = new DOMRect(1, 1, 1, 1);
  C.prototype.__proto__ = (i < 19) ? {} : rect;
  rect.x;      // cache a handler for DOMRect.x
  new C().m(); // reuse it, but invoke on C
}
for (let i = 0; i < 20; i++) main(i);
\end{lstlisting}

\parh{Case Study: CVE-2022-1134.}~We illustrate this limitation with CVE-2022-1134~\cite{cve20221134}, a high severity vulnerability in V8. V8 uses Inline Caches (ICs) to accelerate object property accesses. Line 5 accesses \texttt{rect.x}, causing V8 to cache a handler for the native \texttt{DOMRect.x} accessor. In the first 19 iterations, \texttt{C.prototype.\_\_proto\_\_}, which determines the superclass prototype of \texttt{C}, points to an empty object (line 4), so the cached handler is never invoked when \texttt{super.x} is called on a \texttt{C} instance (line 6). In the final iteration, \texttt{super.x} starts its lookup from the \texttt{DOMRect} object and therefore reuses the cached native accessor handler. However, JavaScript \texttt{super} semantics preserve the current \texttt{C} instance as the accessor receiver. V8 applies a handler specialized for a \texttt{DOMRect} receiver to an incompatible \texttt{C} instance, causing type confusion.

The bug stems from the mismatch between IC handler construction and execution phases. Handler construction checks the \texttt{Map}, which is V8's runtime object-layout descriptor, of the lookup-start object, whereas handler execution invokes the native accessor on \texttt{p->receiver()} without checking that receiver's Map. The cached handler is therefore accepted based on the \texttt{DOMRect} prototype but executed on a \texttt{C} instance.

\begin{lstlisting}[language=js]
// Handler construction: LoadIC::ComputeHandler
Handle<Map> map = lookup_start_object_map();
...
LookupHolderOfExpectedType(map, &holder_lookup);
\end{lstlisting}

\begin{lstlisting}[language=js]
// Handler execution: AccessorAssembler::HandleLoadAccessor
...
CallApiCallback(..., p->receiver());
\end{lstlisting}

\parh{The Missing Feedback Signal.} A fuzzer guided by structural coverage readily covers all the lines above. The vulnerability instead depends on a relationship that edge coverage cannot represent. In the vulnerable execution, the cached handler is compatible with \texttt{lookup\_start\_object} to pass the validation in handler construction, but incompatible with \texttt{receiver} to cause a type confusion in handler execution. Consequently, safe and vulnerable executions can traverse the same control-flow edges while differing in whether the cached handler is compatible with the actual receiver.

\begin{figure*}
    \centering
    \includegraphics[width=0.9\textwidth]{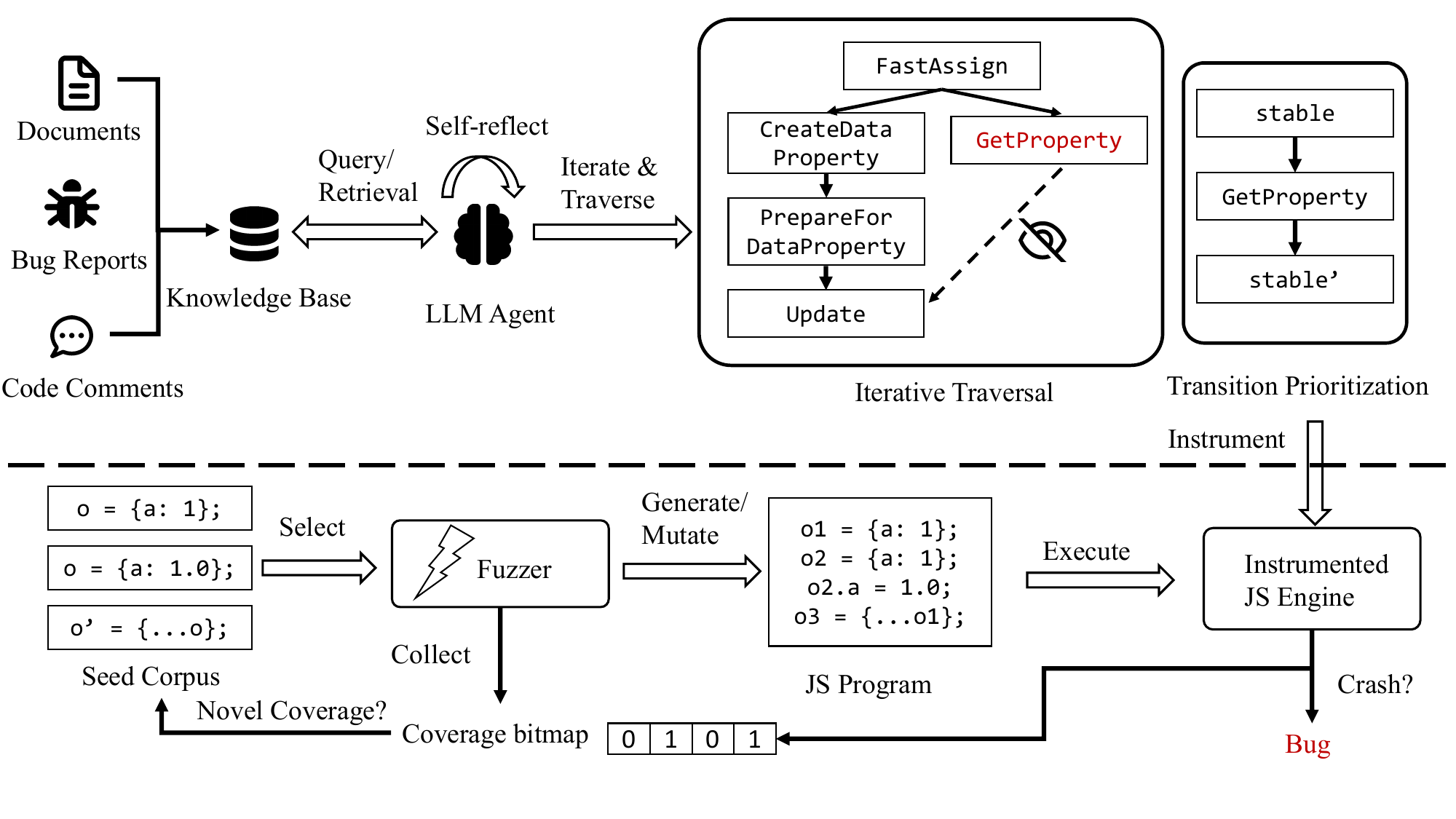}
    \caption{The overall architecture of \tool, which consists of two main phases: (1) an offline analysis and instrumentation phase guided by an LLM agent that extracts semantic beacons from developer artifacts and source code to identify critical states; and (2) an online fuzzing phase where the instrumented engine provides dual feedback to guide input generation toward unexplored semantic states.}
    \label{fig:overview}
\end{figure*}

\subsection{State Coverage as Fuzzing Feedback}
\label{subsec:observations}

To overcome the limitations of structural coverage, state coverage can effectively distinguish executions that traverse the same control-flow edges but differ in internal state.

\parh{Inapplicability of Prior State-Coverage Approaches.} Although prior works have explored state coverage, these approaches are not directly applicable to JS engines. IJON requires developers to manually identify and annotate the states to track~\cite{aschermann2020ijon}, while SDFuzz represents states using call-stack configurations~\cite{Li2024}. Manual annotation does not scale to JavaScript engines, whose millions of lines of code and tightly interacting subsystems expose an enormous number of potential states. Call-stack configurations, meanwhile, capture calling context but not semantic runtime conditions, and therefore cannot distinguish executions that share the same call stack but differ in critical variable values or object configurations.

The core challenge is therefore to automatically and effectively identify JS
engine states to instrument. To understand what states facilitate bug discovery,
we analyzed 35 distinct vulnerabilities from Google's public v8CTF
tracker~\cite{v8ctf-tracker} that were fixed by December 31, 2025. Each entry
includes a working exploit~\cite{v8ctf-rules}. For each case, two authors manually inspected the exploits and corresponding patches and resolved classification disagreements through discussion.
We derive three observations from this study
that inform \tool's state coverage design. These observations are not mutually exclusive.

\parh{O1: Bug-Relevant Conditions Cross Implementation Boundaries.}
In 31 of 35 cases (88.6\%), a condition established in one function or engine
phase was invalidated or consumed in another. In issue
391907159~\cite{v8-391907159}, V8's Wasm code garbage collector marked an import
wrapper as dying, but the wrapper cache reused it before reclamation completed.
The wrapper was then freed while still referenced, causing a use-after-free.

\parh{O2: Side Effects and Internal Transitions Invalidate Assumptions.}
In 12 of 35 cases (34.3\%), a side effect or engine-internal transition changed
state on which other code still relied. These transitions involved object
layout, compiler tier, allocation, garbage collection, and value
representation. In issue 400052777~\cite{v8-400052777}, an operation reached an
object through a different reference and changed its map through an
elements-kind transition. TurboFan retained its earlier layout inference,
causing downstream optimization to produce a type confusion.

\parh{O3: Developer Artifacts Reveal Semantically Important States.} Engine developers record critical state indicators in assertions, typed enumerations, comments, design documents, and bug reports. We call these artifacts \emph{semantic beacons}, which are not itself states, but rather developer-authored evidence that helps \tool\ identify states to instrument. For example, V8 contains more than 33,000 debug assertions and 772 enum class definitions, including engine concepts such as optimization tiers, IC states, and allocation modes. These signals provide scalable starting points for identifying behavior-relevant state dimensions without instrumenting every variable.

\subsection{Combining Semantic Reasoning with Fuzzing}

Automatically extracting these semantic beacons at scale is a complex task. Assertion macros span multiple preprocessor layers, and critical context is frequently embedded in natural language comments rather than machine-readable abstractions. Traditional static analysis tools struggle with this interplay of cross-file dependencies and informal developer intent. Recent advances in applying Large Language Models (LLMs) to program analysis suggest that these artifacts can be mined automatically. Prior work shows that LLMs can infer useful program invariants from code~\cite{iclrinvariants2023} and synthesize semantically meaningful predicates and instrumentation for fuzzing~\cite{zhu2025locus}.

Enabling LLMs for state-aware fuzzing does not come without challenges. Directly applying LLMs to the entire JS engine codebase is infeasible due to the sheer size of the code and the complexity of the interactions between components. Instead, we use an LLM agent for semantic selection, while grounding its reasoning with retrieval and conventional call-graph and data-flow tools. The LLM is used offline for semantic reasoning, while the fuzzer remains responsible for high-throughput exploration and concrete bug triggering.

\section{Overview}
\label{sec:overview}

\cref{fig:overview} shows two stages: offline LLM-guided instrumentation and
online state-aware fuzzing. Offline, the agent
analyzes a knowledge base to synthesize targeted probes for critical internal
states. Online, dual feedback rewards inputs that exercise new semantic states
after edge coverage saturates, guiding the fuzzer toward bugs that conventional
coverage feedback misses.

We define a \emph{state} as a side-effect-free expression over runtime values, which can be variables or side-effect-free functions (observation functions). For an operation with pre- and post-operation program points, the ordered pair of states at those points defines a \emph{state transition}. \emph{State coverage} is the set of previously unseen states or transitions recorded in the semantic bitmap. Two executions contribute different coverage when they produce different observed values at the same site, subject to bitmap hashing.

\subsection{Semantic Beacon Extraction}
\label{subsec:kb}
A \emph{semantic beacon} is itself not a state, but evidence from developer artifacts suggesting a potential critical state or transition to instrument. For example, an assertion may encode an invariant over an object's storage mode, while a bug report may describe a transition that violated such an invariant. Semantic extraction aims to mine these beacons from developer artifacts and map them to relevant states. \tool\ performs this extraction in three steps:

\parh{Knowledge Base Construction.}~Developer artifacts such as design
documents, in-source comments, and historical bug reports record which
internal states matter and which subsystems manage them.
\tool\ aggregates these sources into a vector-indexed \emph{knowledge base}
(KB); \cref{sec:implementation} gives further construction details.

\parh{Knowledge Retrieval.}~The agent consults the KB on demand rather than
loading all artifacts at once. At each discovery step, it issues a targeted
natural-language query and retrieves ranked snippets that connect a semantic
feature to relevant files and functions.

\parh{Self-Reflection Filtering.}~A query can retrieve snippets that use the
right words for the wrong entity or transition. Through self-reflection
prompting~\cite{selfrefine2023}, we instruct the agent to inspect the retrieved snippets, justify why they qualify for the semantic beacon, and output \texttt{keep} or \texttt{reject}.

\parhs{Running Example.}~We use CVE-2024-5830~\cite{cve20245830} as a running example to show how \tool\ performs semantic beacon extraction.

\begin{lstlisting}[language=js, caption={V8's map update and fast-mode assertion.}, label={lst:beacon_example}]
Handle<Map> Map::PrepareForDataProperty(...) {
  map = Update(isolate, map); // replace outdated layouts
  DCHECK(!map->is_dictionary_map()); // expects fast mode
  ...
}
\end{lstlisting}

V8 associates each JS object with an internal layout descriptor, called a \texttt{Map}. Besides storing object properties, the descriptor records whether object properties use a compact array representation (``fast mode'') or a hash-table representation (``dictionary mode''). The code in \cref{lst:beacon_example} prepares such a descriptor before adding a property. Its assertion expresses the assumption that after \texttt{Map::Update} updates an outdated descriptor, the \texttt{Map} should still use fast-mode storage.

A historical bug report~\cite{chromium40062884} documents that \texttt{Update} can return a dictionary map when the hidden-class transition table is exhausted. While such a bug is already fixed, it inspires the agent to generalize the underlying bug pattern by paying attention to transitions across \texttt{Map::Update}, as changing a map from fast to dictionary mode may be security-relevant.

\begin{lstlisting}[style=beaconstyle,escapeinside={(*@}{@*)},label={lst:beacon_retrieval},caption={Phase~1 KB query and representative results.}]
(*@\ding{172}@*) <KB query> "Map::Update map storage mode transition"
<Retrieved snippets>
  [1] comment     an outdated map may cache the map to which
                  objects should migrate.
  [2] comment     walks parent links to test whether maps 
                  belong to the same context.
  [3] design doc  describes the proposed FixedMap collection
\end{lstlisting}

The agent then queries the KB for evidence connecting \texttt{Map::Update} to changes in map storage mode. The query returns three candidates. Candidate [1] explains that an outdated map may cache the map to which objects should migrate, helping establish how \texttt{Map::Update} can replace the caller's original map. Candidate [2] also concerns V8 maps, but only checks whether maps in a transition tree belong to the same native context. Candidate [3] describes the ECMAScript \texttt{FixedMap} collection, where ``map'' refers to a JavaScript collection rather than V8's internal object-layout descriptor.

Only Candidate [1] is relevant to the target transition. Candidate [2] concerns transition-tree membership rather than storage mode, while Candidate [3] refers to a different abstraction altogether. Because keyword overlap alone cannot distinguish these cases, \tool\ prompts the LLM to interpret each snippet against the target fast-to-dictionary transition and justify whether it should be retained, with the results are shown in \cref{lst:beacon_filtering}.

\begin{lstlisting}[style=beaconstyle,escapeinside={(*@}{@*)},label={lst:beacon_filtering},caption={Phase~1 self-reflection: resolving the retrieved snippets against the target storage-mode transition.}]
(*@\ding{173}@*) <Self-reflection>
Target: map storage mode across Update()

  keep   [1] links the old hidden-class map to its
             migration target
  reject [2] checks native-context membership in the
             transition tree, not storage mode
  reject [3] describes an ECMAScript FixedMap,
             not V8's internal hidden class
\end{lstlisting}

The LLM then combines three pieces of evidence. The source assertion shows that \texttt{Map::PrepareForDataProperty} expects the result of \texttt{Map::Update} to remain in fast mode; the historical bug report shows that this expectation can be violated; and Candidate [1] explains how an outdated map can be replaced through the migration path. Together, these sources yield the beacon summarized in step~\ding{174} of \cref{lst:beacon_summary}.

\begin{lstlisting}[style=beaconstyle,escapeinside={(*@}{@*)},label={lst:beacon_summary},caption={Phase~1 Beacon Summary for the running example.}]
(*@\ding{174}@*) <Beacon Summary>
  state descriptions: map deprecation, storage mode
  transition hint  : deprecated fast map
                     -> dictionary map via Map::Update
  seed symbols     : Map::PrepareForDataProperty,
                     Map::Update
\end{lstlisting}

At this point, \tool\ has identified the relevant states using natural language descriptions, the suspected transition, and the seed functions from which to continue analysis. Phase 2 starts from these symbols and traces the interprocedural path by which a deprecated fast map can be updated into a dictionary map.

\subsection{Phase~2: Iterative State Discovery}
\label{subsec:state-discovery}

Phase 2 builds on semantic beacon extraction by expanding seed source symbols to track where a target state is established, modified, and consumed. This interprocedural tracing is based on the observation O1 in \cref{subsec:observations} that bug-relevant conditions often span multiple functions and are rarely visible from a single source location.

To accomplish this, the agent iteratively explores a frontier of functions and variables, guided by the Beacon Summary's state descriptions and transition hints. It utilizes call-graph traversal, data-flow analysis, source inspection, and KB retrieval to uncover hidden semantic contexts and track states across functions.

\parhs{Running Example.}~The agent expands the fully qualified
seed symbols in the Beacon Summary. Because C++ function names can be reused,
it checks each call-graph candidate against the source and retains only edges
that invoke the target function. For the running example, the validated edges
form the two branches shown in step~\ding{172} of
\cref{lst:guided_traversal}.

\begin{lstlisting}[style=beaconstyle,escapeinside={(*@}{@*)},label={lst:guided_traversal},caption={Evidence-guided call-graph expansion and query refinement.}]
(*@\ding{172}@*) CallGraph(Map::Update)

source-validated direct callers
  Map::PrepareForDataProperty
  JSObject::MigrateInstance

source-validated caller chain from
  Map::PrepareForDataProperty
  <- TryFastAddDataProperty
  <- CreateDataProperty
  <- FastAssign

The migration branch supplies the next query

  (*@\ding{173}@*) <KB query> "JSObject MigrateInstance
     deprecated map migration"
\end{lstlisting}

\parh{Query Refinement.}~The migration branch contributes the
\texttt{JSObject::\allowbreak MigrateInstance}. The agent combines this symbol
with the beacon's deprecated map context in the refined query shown in
step~\ding{173} of \cref{lst:guided_traversal}. Later searches can then focus
on the migration code that changes map storage.

\begin{lstlisting}[style=beaconstyle,escapeinside={(*@}{@*)},label={lst:state_report},caption={State Report for the running example.}]
(*@\ding{174}@*) <State Report>
WHAT : whether Map::Update converts a deprecated
       fast map into a dictionary map.
WHY  : TryFastAddDataProperty later uses the returned
       map in descriptor and write logic that assumes
       fast-property storage.
HOW  : an accessor can deprecate the target transition
       map before CreateDataProperty reuses it; updating
       that map may produce dictionary-mode storage.
Candidate sites
       immediately before and after Map::Update in
       Map::PrepareForDataProperty
\end{lstlisting}

\parh{From Traversal to a State Report.}~The two branches meet at
\texttt{Map::Update}. The migration branch follows
\texttt{JSObject::\allowbreak MigrateInstance} into the deprecated map update
path. The caller branch ascends from
\texttt{Map::\allowbreak PrepareForDataProperty} via
\texttt{TryFastAdd\allowbreak DataProperty} and \texttt{CreateDataProperty} to
\texttt{FastAssign}; it exposes the downstream fast-map use and the accessor
that establishes the transition precondition. For this example, the branches
ground the state-changing operation, its downstream use, and their concrete
source locations. A partial report produced at the step limit still undergoes
self-reflection; reports with no surviving locations are discarded. Step~\ding{174} in \cref{lst:state_report} shows the resulting report.

\begin{lstlisting}[language=js, caption={Simplified excerpt of V8's \texttt{FastAssign} (\texttt{js-objects.cc}). The \texttt{stable} flag tracks source-map changes during property reads; the bug-relevant update occurs later on the target.}, label={lst:fastassign}]
bool stable = true;
for (InternalIndex i : map->IterateOwnDescriptors()) {
  if (stable) {
    // Fast path: decode directly from descriptor array
    if (details.kind() == kData) {
      prop_value = FastPropertyAt(from, ...);
    } else {
      // Getter: may trigger side effects
      prop_value = Object::GetProperty(&it);
      stable = from->map() == *map;  // map changed?
    }
  } else {
    // Slow path: fresh lookup
    prop_value = Object::GetProperty(&it);
  }
  CreateDataProperty(target, next_key, prop_value);
}
\end{lstlisting}

\parh{Causal Chain.}~The caller branch reaches \texttt{FastAssign}
(\cref{lst:fastassign}). Before cloning, a transition map already exists for
adding the copied property. \texttt{FastAssign} evaluates an accessor before
calling \texttt{CreateDataProperty}. By creating a conflicting field
representation, the accessor deprecates that transition map.
\texttt{TryFastAdd\allowbreak DataProperty} then reuses the deprecated map and
passes it to \texttt{Map::\allowbreak PrepareForDataProperty}.
\texttt{Map::Update} may return a dictionary map, yet the following
\texttt{WriteToField} still assumes fast-property storage, causing the type
confusion~\cite{cve20245830}.

\subsection{Phase~3: Transition Prioritization and Probe Synthesis}
\label{subsec:transition-prioritization}

\parh{Side-Effect-Driven Transitions.}~O2 in
\cref{subsec:observations} motivates \tool\ to prioritize state changes
caused by side effects. In \texttt{FastAssign}, an accessor can
deprecate the target transition map before \texttt{CreateDataProperty} reuses
it. Edge coverage follows the same path whether \texttt{Map::Update} returns a
fast or dictionary map.

\parh{Selecting Relevant Transitions.}~\tool\ follows call-graph and data-flow
links from each report, then selects in-scope expressions that observe the
reported state without side effects.

\parhs{Running Example.}~The analysis exposes three
candidates: \texttt{stable}, \texttt{is\_deprecated()}, and
\texttt{is\_dictionary\_map()}. The first directly represents source-map
stability, but it does not directly reflect the selected target-map property;
\tool\ therefore rejects it. The other two expressions capture the target
map's transition precondition and downstream outcome.
Recording storage mode on both sides of \texttt{Map::Update} further
distinguishes the bug-relevant fast-to-dictionary update from a map already in
dictionary mode. This combination of states captures the update and
the later fast-storage assumption. Step~\ding{172} in
\cref{lst:state_expr} shows the selected expressions as states to instrument; step~\ding{173} shows
the synthesized probe.

\begin{lstlisting}[
  style=beaconstyle,
  escapeinside={(*@}{@*)},
  label={lst:state_expr},
  caption={Selected states and the synthesized probe.}
]
(*@\ding{172}@*) Selected transition
  [src/objects/map.cc, <before Update>,
   (*@\texttt{\textless}@*)map->is_deprecated(),
    map->is_dictionary_map()(*@\texttt{\textgreater}@*)]
  [src/objects/map.cc, <after Update>,
   map->is_dictionary_map()]

(*@\ding{173}@*) Synthesized probe
  bool deprecated_before = map->is_deprecated();
  bool dictionary_before = map->is_dictionary_map();
  map = Update(isolate, map);
  uint32_t pre_state =
      (deprecated_before (*@\texttt{\textless\kern0pt\textless}@*) 1) | dictionary_before;
  SEMANTIC_CONTEXT_ENUM(id, pre_state,
                   map->is_dictionary_map());
\end{lstlisting}

The probe packs the two pre-state predicates into one value and pairs it
with the post-update storage mode. The resulting bitmap entry distinguishes the
deprecated-fast-to-dictionary transition from other outcomes of
\texttt{Map::Update}.

\subsection{Instrumentation Stage}
\label{subsec:instrumentation}

Phase~3 outputs the source locations and state expressions
to instrument. The instrumentation stage translates these specifications into
code patches that expose the selected transitions through lightweight probes.

\parh{Instrumentation Design.}~As mentioned in \cref{subsec:transition-prioritization}, critical feedback signals often arise from state transitions. To capture these transitions, \tool\ inserts temporary buffers into the engine that record pre-transition values of the relevant state expressions, then hashes the combination of the pre- and post-transition values into a shared-memory bitmap that the fuzzer can read. This design allows the fuzzer to recognize when a critical state transition occurs, even if the control flow remains unchanged.

\parh{Instrumentation and Memory Management.}~Given the selected state
expressions and their source locations, probe insertion first adds temporary
buffers for pre-transition values and then emits probe calls based on each
observed value's type. At JS engine startup, a dedicated initialization routine
creates or opens a POSIX shared memory region keyed to the process ID, with a
fixed layout comprising the bitmap and the maximization slot array (for
tracking integer states). This region is mapped into the engine's address space
and into the fuzzer's address space, enabling zero-copy state transfer. The
region is zeroed at initialization and remains writable throughout the
execution lifetime.

\subsection{Dual-Feedback Fuzzing}
\label{subsec:dual-feedback}

During fuzzing, generated JS programs are executed on the instrumented engine, and those that trigger novel bitmap entries are retained for further mutation. We use a dual-feedback design with two instrumented JS engine instances that share the same seed corpus: one instrumented for structural coverage only, and one instrumented for both structural and state coverage. The fuzzer dynamically switches between them based on structural coverage growth, leveraging the strengths of both signals at different stages of the search process while avoiding state-instrumentation overhead when it is not yet useful.

The structural-only instance is cheaper to execute and is therefore used for early exploration. This lets the fuzzer quickly explore new code paths with dense and informative signal while the input space remains largely unexplored. The fuzzer continuously monitors structural coverage growth and, once it plateaus for long enough (details in \cref{sec:implementation}), switches to the state-augmented instance. This enables it to use finer-grained state signal to guide the search toward inputs that trigger specific state transitions, which are often necessary to expose deep bugs invisible to structural coverage.

\section{Implementation of \tool}
\label{sec:implementation}

\parh{Instrumentation and Feedback.}~We compile the instrumented engine to identify syntax, type, and scope errors, run its existing test suite to detect unintended side effects such as incorrect reference-count updates, and prompt the LLM to repair whatever either step reports.

We run the fuzzer and the target engine as separate processes, so the fuzzer cannot perturb the engine's internal state and a crashing or hanging engine cannot disrupt the fuzzing service.
To reduce the overhead of inter-process communication (IPC), we implement a shared-memory region that the target engine can write to when an instrumentation probe is triggered, and the fuzzer can read from it after each test execution for feedback collection. We implement a small set of probe utility functions that the instrumentation can call to write to the shared-memory bitmap. For example, \texttt{sem\_set} takes an index and sets the corresponding bit in the bitmap. By design, these functions are read-only with respect to engine state and write exclusively to the shared-memory bitmap, and thus shall not cause side effects on the engine's normal execution.

\parh{Integration with Existing Fuzzers.}~We implement \tool\ on top of
Fuzzilli~\cite{gross2023fuzzilli}, a coverage-guided fuzzer whose modular,
multi-engine design supports our evaluation across different targets.
Within Fuzzilli, we switch to state-augmented feedback after 100 consecutive
iterations yield no new structural edge; discovering an edge resets the
counter. We use this threshold for every engine and run.
In total, we wrote 397 lines of C to support the
instrumentation and feedback infrastructure, and 2774 lines of
Swift to implement the IPC channel between the fuzzing module and the
target engine binary, enabling collection of runtime feedback and
propagation of state information back to the fuzzer.

\parh{Knowledge Base Construction.}~To construct the knowledge base in
\cref{subsec:kb}, we crawl three source types for each engine besides source
code: (1) security-labeled reports from the engine's bug tracker, (2) in-source
comments extracted via AST traversal of the engine's source code repository, and
(3) developer design documents such as ECMAScript specifications and
engine-specific design docs. Documents are chunked and indexed into a vector
store for retrieval-augmented queries. 

\parh{Program Analysis Tools.}~In order to enable efficient navigation of the codebase, we implement a set of program analysis tools that the agent can use to query the codebase. Besides simple file search and text grep, we implement three tools for interprocedural call-graph traversal, def-use tracking, and structural pattern matching. These tools are listed in \cref{tab:agent-tools}:
(1) \textsc{CallGraph} tool enables the agent to perform interprocedural call-chain traversal, which facilitates tracing the flow of states across function boundaries. (2) \textsc{DataFlow} tool allows the agent to track the definitions and uses of variables, enabling it to identify how states are defined and propagated through the code. (3) \textsc{ASTMatch} tool enables the agent to perform structural pattern matching across the codebase, which can help identify code regions that manipulate states via certain patterns. To bound the exploration cost, we cap each semantic beacon at 25 tool-using steps. Nonetheless, these tools are not the only means for the agent to navigate the codebase; it can also leverage the KB to find relevant code snippets. We also admit that while more advanced program analysis tools such as symbolic execution could potentially be helpful for the agent to understand the codebase, they are not currently implemented in our prototype due to engineering complexity and potential scalability issues. We leave the integration of more advanced program analysis tools as future work.

\begin{table}[t]
\caption{Code analysis tools available to the agent.}
\label{tab:agent-tools}
\centering
\small
\begin{tabular}{llll}
\toprule
\textbf{Tool} & \textbf{Input} & \textbf{Output} \\
\midrule
\textsc{CallGraph} & Function name & Callers and callees \\
\textsc{DataFlow} & Variable, scope & Def-use sites \\
\textsc{ASTMatch} & Structural pattern & Matching AST nodes \\
\bottomrule
\end{tabular}
\end{table}

\section{Evaluation}
\label{sec:evaluation}

We evaluate \tool\ on six JavaScript engines to answer three questions:

\begin{itemize}[leftmargin=*]
  \item \textbf{Q1:} Can \tool\ find real-world bugs in JavaScript engines? (\cref{subsec:bug-finding})
  \item \textbf{Q2:} How does \tool\ compare to state-of-the-art JS engine fuzzers? (\cref{subsec:comparison})
  \item \textbf{Q3:} How do individual components contribute to effectiveness? (\cref{subsec:ablation})
\end{itemize}

\subsection{Evaluation Setup}
\label{subsec:evaluation-setup}

\parh{JS Engines Under Test.}~We evaluate \tool\ on six JS engines with different codebase sizes and use case scenarios:
V8, SpiderMonkey, JavaScriptCore, QuickJS, Hermes, and Escargot.
These engines serve as critical infrastructure beyond just browser runtimes:
V8 drives edge compute for major cloud
providers~\cite{cloudflareworkers},
SpiderMonkey and QuickJS are embedded in databases such as
MongoDB~\cite{mongodb} and CouchDB~\cite{couchdb},
JavaScriptCore underpins server-side runtimes like
Bun~\cite{bunjsc},
Hermes is the default engine for React
Native~\cite{reactnative}, and
Escargot powers system services on Samsung's Tizen OS for IoT
devices.

All the six engines are mature, widely used, and have been
continuously stress-tested in the wild, with extensive fuzzing efforts from both internal teams (using tools like OSS-Fuzz and in-house fuzzers) and the security community. Given this, they represent a challenging testbed for evaluating \tool's bug-finding capabilities, as many of the low-hanging fruits have already been discovered and fixed.

\parh{Experimental Environment.}~
All experiments run on an Ubuntu 22.04~LTS workstation with a
64-core Ryzen 9980X processor and 128\,GB of RAM. We compile each
engine with LLVM~22~\cite{lattner2004llvm} and enable
AddressSanitizer~\cite{serebryany2012addresssanitizer}. Every crashing
input is replayed on the same engine revision and sanitizer
configuration without \tool\ probes, and we count only failures that
reproduce.
Probe synthesis uses GPT-5.1~\cite{gpt51} at temperature zero to reduce
sampling variance.

\parh{Instrumentation Cost.}~
One end-to-end instrumentation synthesis takes 19--80 minutes, with
an average of 48 minutes across the six engines. QuickJS completes
fastest, while V8 takes longest because of its larger codebase. Each
synthesis consumes an average of 1.02M input tokens and 0.51M output
tokens per engine. The average monetary cost is \$6.45, comprising
LLM calls (\$3.20), retrieval (\$1.50), re-instrumentation (\$1.00),
and validation-driven repair (\$0.75). The per-engine cost ranges
from \$0.28 for QuickJS to \$13.75 for V8 and broadly follows
codebase size.

\parh{Runtime and Memory Overhead.}~
We measure runtime overhead by comparing the end-to-end throughput of Fuzzilli
and \tool\ on V8 over a 24-hour fuzzing session.
Fuzzilli executes 9.35M test cases and \tool\ 8.83M, a 5.5\% runtime overhead.

For memory, we sample resident set size (RSS) across all 10 QuickJS instances
for five minutes after 24 hours of fuzzing. Mean RSS is 23\% higher with
\tool\ probes than without them.

\subsection{Bug Finding}
\label{subsec:bug-finding}

\parh{Overall Effectiveness.}~During a three-month continuous fuzzing
campaign, \tool\ uncovered 68 bugs across six JavaScript engines: 25 in V8,
14 in JavaScriptCore, 9 in Escargot, 7 each in SpiderMonkey and QuickJS, and 6
in Hermes. Throughout the campaign, we periodically updated each engine to
its latest release and reinstrumented each new revision, so subsequent runs
targeted current code rather than bugs already fixed upstream. These engines
have been extensively audited by their vendors, stress-tested by
OSS-Fuzz~\cite{ossfuzz}, and covered by vendor bug bounty programs~\cite{chromevrp}; the
fact that \tool\ still finds new bugs in such mature codebases underscores
its effectiveness.

\textit{Independent Root Causes.}~The 68 bugs represent independent root
causes: upstream maintainers tracked them in separate tickets and fixed them
with distinct patches. These causes span execution tiers and compiler
subsystems, including graph construction, IR lowering, register allocation,
deoptimization dispatch, and the interaction between JIT constant emission
and shared-heap garbage collection.

\textit{Security Impact.}~At least 35 bugs were confirmed to have security
implications, including remote code execution and sandbox escapes. The V8
security team and Meta awarded bounties for several discoveries, Apple credited
us in security updates, and some bugs were adopted as capture-the-flag (CTF)
challenges~\cite{hkcertctf}.

\begin{lstlisting}[language=js, caption={Minimized QuickJS PoC.},
  label={lst:qjspoc}]
(function(){
  let w = new os.Worker('aaaa');
  w.onmessage = function(){};
  Object.defineProperty(w, 'onmessage', {
    set(v){
      delete w.onmessage;
      w.postMessage = null;
    }
  });
  w.onmessage = 42;  // triggers setter
})();
\end{lstlisting}

\parh{Case Study.}
We show two of our uncovered bugs in detail to illustrate why state-sensitive probes are crucial for finding them. These bugs are only discovered by \tool\ and are missed by all the vendor fuzzers and security community. In fact, these bugs have been present in the codebase for years and have been continuously fuzzed by the respective engine developers, yet they remain undetected until \tool\ found them.

\parh{QuickJS: Use-After-Free in Worker.}~
\cref{lst:qjspoc} shows the PoC for a use-after-free bug in
QuickJS's Worker implementation.
Line~2 creates a Worker, which registers an internal communication
port in the runtime's thread state.
Lines~3--9 install an ordinary \texttt{onmessage} handler and then redefine
\texttt{onmessage} as an accessor whose setter deletes that handler and
nullifies the Worker's \texttt{postMessage} reference.
Line~10 assigns to \texttt{onmessage}. Because the property is now an accessor,
the assignment runs the setter and
leaves the Worker's internal port references out of sync with the
runtime's port list.
When the program exits, the runtime first frees its thread state, which holds
the head of the port list, and then runs a final GC pass that invokes the
Worker's finalizer.
Because the port was never unlinked, the finalizer reaches
\texttt{js\_free\_port} (\cref{lst:qjsprobe2}), whose \texttt{list\_del} writes
through two neighbor pointers that still reference the freed thread state,
producing a use-after-free write.

Two probes expose the state combinations behind this bug. Both use
\texttt{SEMANTIC\_CONTEXT\_ENUM}, which takes a site identifier and two engine
values and sets the bitmap entry for that triple, so the fuzzer distinguishes
\emph{combinations} of engine values at a site rather than individual values.
The probe in \cref{lst:qjsprobe1} records the old and new property types at
every \texttt{defineProperty} redefinition; in the PoC it fires on the
operation that converts \texttt{onmessage} from a data property into the
destructive setter.

\begin{lstlisting}[language=js, caption={Probe~1: property type transition in defineProperty (quickjs.c).},
  label={lst:qjsprobe1}]
int JS_DefineProperty(JSContext *ctx, ...) {
  ...
  // old type x new type
  SEMANTIC_CONTEXT_ENUM(0x3144,
    (old_flags & JS_PROP_TMASK) >> 4,
    (flags & JS_PROP_TMASK) >> 4);
  ...           
}
\end{lstlisting}

\cref{lst:qjsprobe2} records two state dimensions when the Worker's
finalizer runs: the runtime's GC phase and the type of the port's
message handler.
When a Worker is reclaimed through the normal cleanup path, the
handler is cleared before GC collects the object, so the finalizer
sees a null handler outside of any GC phase.
The PoC preserves a live handler that persists into the GC cycle,
stressing a finalizer path that the runtime's cleanup ordering does
not anticipate.

A coverage-guided fuzzer quickly exercises these code paths, but edge coverage fails to distinguish the specific state combinations that trigger the bug. Without state-sensitive probes, the fuzzer misses the unusual setup (installing a destructive setter on a Worker property) and the vulnerable cleanup condition (a Worker entering finalizer execution with desynchronized internal references). By making these runtime states explicitly visible, our probes allow the fuzzer to detect and retain the novel combinations that lead to the crash.

\begin{lstlisting}[language=js, caption={Probe~2: Worker state at finalization (quickjs-libc.c).},
  label={lst:qjsprobe2}]
static void js_worker_finalizer(JSRuntime *rt,
    JSValue val) {
  JSWorkerData *w = JS_GetOpaque(val, ...);
  if (w) {
    // gc phase x handler state
    SEMANTIC_CONTEXT_ENUM(0x4016,
      (uint32_t)(rt->gc_phase),
      w->msg_handler
        ? JS_VALUE_GET_TAG(
            w->msg_handler->on_message_func)
        : JS_TAG_NULL);
    js_free_port(rt, w->msg_handler);
  }
}
static void js_free_port(JSRuntime *rt,
    JSWorkerMessageHandler *port) {
  ...
  list_del(&port->link);  // neighbours live in
  js_free_rt(rt, port);   // the freed thread state
}
\end{lstlisting}

\parh{V8: Stale Addresses in the JIT Compiler.}~Unlike the QuickJS bug, which is related to the lifecycle of a single object, this V8 bug involves coordination across multiple subsystems. It was awarded a bug bounty by the V8 security team.

\cref{lst:v8poc} is a simplified version of the PoC. Lines 3--5 force a concatenated string into a globally shared pool by using it as a property key, while line 6 triggers JIT compilation. The vulnerability is triggered in lines 10--11: a Worker thread executes the JIT-compiled code while the main thread invokes a GC in the shared heap that relocates the string. Because this GC only updates the main thread's relocation counter, the Worker's compiler is unaware of the move and operates on a stale memory address, leading to a memory safety violation.

\begin{lstlisting}[language=js, caption={Simplified V8 PoC.},
  label={lst:v8poc}]
function entry() {
  function f() {
    let s = "codePointAt";
    s += "toStringTag";  // stored in shared string pool
    ("h")[s];            // used as property key
    for (let i=0; i<1e6; i++) {} // trigger JIT
  }
  for (let i = 0; i < 100; i++) f();
}
new Worker(entry, {type:"function"});
gc();  // shared-heap GC relocates the string
\end{lstlisting}

This vulnerability highlights the complex interaction between the JIT compiler and garbage collector. Triggering it requires two specific events, which our tool captures using targeted probes:

\begin{lstlisting}[language=js, caption={Probe~1: compiled constant type and location (js-graph.cc).},
  label={lst:v8probe1}]
Node* JSGraph::HeapConstantNoHole(
    Handle<HeapObject> value) {
  SEMANTIC_CONTEXT_ENUM(0x23080,
    (uint32_t)(value->map()->instance_type()),
    (uint32_t)(MemoryChunk::FromHeapObject(
      *value)->owner_identity()));
  ...         // object type x memory region
}
\end{lstlisting}

\cref{lst:v8probe1} is placed where the JIT compiler records a heap
object as a constant in the compiled code. The probe hashes the object's type with the memory region it resides in.
Most compiled constants are read-only objects (maps, builtins) that
the GC never moves; when the compiler embeds a string that lives in
the shared heap, the probe triggers a novel type-region combination that the fuzzer has not previously seen, thus retaining the input for further mutation.

Novel type-region combination alone is not sufficient to trigger the bug; the GC must also perform a compacting collection that can relocate the object. \tool's second probe (\cref{lst:v8probe2}) hashes the GC collector type together with the
collection reason. Most inputs trigger only minor, non-relocating collections; inputs triggering a full mark-compact collection trigger a novel state.

\begin{lstlisting}[language=js, caption={Probe~2: GC collector and reason (gc-tracer.cc).},
  label={lst:v8probe2}]
RecordGCPhasesInfo(Heap* heap,
    GarbageCollector collector,
    GarbageCollectionReason reason) {
  SEMANTIC_CONTEXT_ENUM(0x31023,
    (uint32_t)(collector),
    (uint32_t)(reason)); // collector x reason
  ...
}
\end{lstlisting}

As such, by retaining inputs that cause both probes to hit novel states and mutating them further, \tool\ can find the specific combination of a shared-space constant and a compacting GC that leads to the vulnerability.
An edge-coverage fuzzer cannot distinguish \emph{where} the compiler embeds objects or \emph{which} strategy GC runs, thus cannot guide the input search towards the error-triggering combination.

\subsection{Comparison with Baselines}
\label{subsec:comparison}

\parh{Baselines.}~
We compare \tool\ against four state-of-the-art JavaScript engine
fuzzers.
Since \tool\ is built on top of Fuzzilli~\cite{gross2023fuzzilli},
an IR-based generation fuzzer, the Fuzzilli comparison serves as a
controlled ablation: the two tools share the same mutation engine,
corpus generation strategy, and runtime infrastructure, differing only
in \tool's state-sensitive feedback.
The remaining baselines are
DIE~\cite{park2020fuzzing}, an aspect-preserving mutation fuzzer;
OptFuzz~\cite{wang2024optfuzz}, a JIT optimization-path-guided fuzzer;
and HLPFuzz~\cite{hlpfuzz}, an LLM-based constraint-solving fuzzer.
OptFuzz targets JIT-specific behavior and is therefore
evaluated only on the three JIT-enabled engines (V8, JSC,
SpiderMonkey).
We use the latest versions of each fuzzer at the time of evaluation
with their default or recommended configurations.
For fuzzers that require seed inputs, following
prior work~\cite{park2020fuzzing}, we use the test suites shipped
from each engine's repository as the initial corpus for fuzzing that engine.
Note that \tool\ does not require seed inputs.

\parh{Setup.}~
To account for randomness, we evaluate each fuzzer-engine combination 5 times independently. Each of these runs executes the entire pipeline from scratch, including re-instrumentation of the target engine.
Following common evaluation practices~\cite{wang2024optfuzz},
each run uses 10 dedicated CPU cores and lasts for 72 hours.

\parh{Bug Comparison.}~
We use bug-finding as the primary evaluation criterion, as it is an
objective end-to-end measure independent of any coverage metric.
We deduplicate crashes by manual root-cause analysis and report
unique bugs per engine.
Over the 72-hour evaluation window, \tool\ discovers 39 unique bugs
in total, while the second-best fuzzer, Fuzzilli, only finds 23 (a 70\%$\uparrow$).
Notably, 14 bugs found by \tool\ are not triggered by any
baseline.

\parh{Coverage Comparison.}~We replay every fuzzer's final corpus on two common builds of each engine.
The standard edge-coverage build contains no \tool\ probes. The second build
contains the same fixed probe set for every corpus and records unique entries
in the shared-memory bitmap (\cref{subsec:instrumentation}). We normalize
state coverage per engine to Fuzzilli, whose coverage is 1.0$\times$.

As shown in \cref{tab:comparison}, all fuzzers achieve comparable edge
coverage. \tool\ ranks first on three of the six engines, and its absolute gap
from the best baseline remains below one percentage point on every engine.
State coverage separates the fuzzers more clearly: DIE, OptFuzz, and HLPFuzz
reach 1.01$\times$, 1.23$\times$, and 1.18$\times$ Fuzzilli's state coverage
on average, whereas \tool\ reaches 1.72$\times$. The baselines therefore
exercise fewer states despite reaching similar code. 
The 14 bugs found exclusively by \tool\ require this state-coverage feedback.
Even when edge coverage saturates, as in QuickJS, \tool\ continues to
trigger new semantic-bitmap entries corresponding to necessary vulnerability
conditions such as those in \cref{subsec:bug-finding}.

\begin{table*}[t]
\centering
\small
\caption{Coverage over 72h (median of 5 end-to-end runs).
Edge: percentage of edges covered.
State: semantic-bitmap entries relative to Fuzzilli (1.00$\times$).
Best result per engine is \textbf{bolded}.
``N/A'' indicates the fuzzer does not support that engine.}
\label{tab:comparison}
\begin{tabular}{l rr rr rr rr rr rr}
\toprule
& \multicolumn{2}{c}{\textbf{V8}}
  & \multicolumn{2}{c}{\textbf{JSC}}
  & \multicolumn{2}{c}{\textbf{SM}}
  & \multicolumn{2}{c}{\textbf{QJS}}
  & \multicolumn{2}{c}{\textbf{Hermes}}
  & \multicolumn{2}{c}{\textbf{Escargot}} \\
\cmidrule(lr){2-3} \cmidrule(lr){4-5} \cmidrule(lr){6-7}
\cmidrule(lr){8-9} \cmidrule(lr){10-11} \cmidrule(lr){12-13}
\textbf{Fuzzer}
  & \textbf{Edge} & \textbf{State}
  & \textbf{Edge} & \textbf{State}
  & \textbf{Edge} & \textbf{State}
  & \textbf{Edge} & \textbf{State}
  & \textbf{Edge} & \textbf{State}
  & \textbf{Edge} & \textbf{State} \\
\midrule
Fuzzilli
  & 15.30\% & 1.00
  & 24.45\% & 1.00
  & \textbf{26.54\%} & 1.00
  & 51.60\% & 1.00
  & 25.37\% & 1.00
  & 42.03\% & 1.00 \\
DIE
  & 13.12\% & 1.15
  & 21.75\% & 1.04
  & 20.39\% & 0.74
  & 50.91\% & 1.03
  & 24.52\% & 0.95
  & 39.85\% & 1.17 \\
OptFuzz
  & 14.70\% & 1.48
  & \textbf{25.84\%} & 1.18
  & 25.58\% & 1.04
  & N/A & N/A
  & N/A & N/A
  & N/A & N/A \\
HLPFuzz
  & 14.24\% & 1.21
  & 24.56\% & 1.13
  & 23.51\% & 1.02
  & 51.92\% & 1.10
  & \textbf{26.50\%} & 1.36
  & 42.22\% & 1.27 \\
\midrule
\tool
  & \textbf{15.85\%} & \textbf{2.04}
  & 24.92\% & \textbf{2.19}
  & 26.13\% & \textbf{1.33}
  & \textbf{52.14\%} & \textbf{1.30}
  & 26.17\% & \textbf{1.75}
  & \textbf{42.88\%} & \textbf{1.68} \\
\bottomrule
\end{tabular}
\end{table*}

\subsection{Component Effectiveness}
\label{subsec:ablation}

To understand the contribution of individual design choices, we
conduct an ablation study by systematically varying three dimensions
of \tool: the language model used in the analysis, the number of inserted probes,
and existence of the knowledge base. We evaluate on V8 and QuickJS, as these are two representative engines with distinct complexity and common use cases (server vs. embedded).
All variants are executed under the same setup as in \cref{subsec:comparison}.
\cref{tab:ablation} reports the results.

\begin{table}[t]
\centering
\caption{Ablation study over 72 hours (median of five independent runs).
Edge: percentage of edges covered.
Bugs: unique bugs found.}
\label{tab:ablation}
\resizebox{0.8\columnwidth}{!}{%
\begin{tabular}{l rr rr}
\toprule
& \multicolumn{2}{c}{\textbf{V8}}
  & \multicolumn{2}{c}{\textbf{QuickJS}} \\
\cmidrule(lr){2-3} \cmidrule(lr){4-5}
\textbf{Variant}
  & \textbf{Edge} & \textbf{Bugs}
  & \textbf{Edge} & \textbf{Bugs} \\
\midrule
\tool\ (full)
  & 15.85\% & 14
  & 52.14\% & 5 \\
\midrule
\multicolumn{5}{l}{\emph{Language model}} \\
\quad w/ GPT-5 Mini
  & 15.64\% & 11
  & 52.01\% & 2 \\
\quad w/ GLM 4.7
  & 15.70\% & 12
  & 51.95\% & 4 \\
\midrule
\multicolumn{5}{l}{\emph{Probe density}} \\
\quad 50\% probes
  & 15.83\% & 10
  & 52.34\% & 3 \\
\quad 30\% probes
  & 15.78\% & 10
  & 51.93\% & 2 \\
\quad 10\% probes
  & 15.74\% & 9
  & 51.82\% & 2 \\
\midrule
\multicolumn{5}{l}{\emph{Knowledge base}} \\
\quad w/o KB
  & 15.79\% & 10
  & 52.14\% & 3 \\
\bottomrule
\end{tabular}%
}
\end{table}
\vspace{-8pt}

\parh{Language Model.}~
To assess sensitivity to the underlying model, we replace GPT-5.1
with two alternatives: GPT-5 Mini~\cite{gpt5mini} and GLM 4.7~\cite{glm47} and
re-run the whole pipeline to generate probes and fuzz with them. GPT-5.1
generates 1,194 probes on V8 and 289 on QuickJS, compared with 872/233 for
GPT-5 Mini and 1,232/346 for GLM 4.7. GPT-5 Mini and GLM 4.7 uncover 13 (11 on
V8, 2 on QuickJS) and 16 (12 on V8, 4 on QuickJS) bugs, respectively, compared
with GPT-5.1's 19 (14 on V8, 5 on QuickJS).

Manual inspection of the generated probes reveals a qualitative difference. GPT-5.1 generates probes that target specific engine-internal semantics, isolating state transitions crucial for bug exposure, such as the exact type assumptions of JIT guards or object statuses at GC safepoints. Conversely, weaker models produce probes that capture only coarse-grained states, like function entry points or high-level branch conditions, and do not correlate with internal state transitions. Although all models generate compilable instrumentation, the disparity in bug discovery can stem from how the probes monitor relevant state changes.

\parh{Probe Density.}~
To measure how the number of probes affects bug finding, we randomly
retain 50\%, 30\%, and 10\% of the original probes and re-evaluate fuzzing performance. As the probe count decreases, the number of bugs found also drops: At 50\% retention, \tool\ finds 10 bugs on V8 and 3 on QuickJS, a reduction of 29\% and 40\% respectively compared to the full probe set.
At 30\%, the counts drop further to 10 and 2.
At 10\%, only 9 and 2 bugs remain, a 42\% overall reduction.
Edge coverage, by contrast, remains stable across all variants, with less than 5\% variation on V8.

\parh{Knowledge Base.}~To evaluate the contribution of the knowledge base, we disable the agent's access to the knowledge base and only provide the source code as context for probe generation. Disabling the knowledge base reduces probe counts by 22\% on V8 and 34.5\% on QuickJS, causing found bugs to decrease from 14 to 10 on V8 and from 5 to 3 on QuickJS. Manual inspection reveals this decline is qualitative as well: KB-free probes target shallow states like temporary variables and explicit branch conditions, while the full pipeline identifies state transitions and cross-component interactions generalized from historical bug patterns. The knowledge base thus steers the agent toward deeper state dimensions that are more likely to expose bugs.

\section{Discussion}
\label{sec:discussion}

\parh{Testing vs. Verification.}~\tool\ adopts the testing approach to enhance the security and reliability of JS engines. Another potential approach is to formally verify the correctness of JS engines. \tool, like any other testing approach, cannot guarantee the absence of bugs. While formal verification offers the promise of mathematically proving software correctness, fully verifying all aspects of JS engines is still an open research problem due to their complexity (millions of lines of code with intricate semantics and optimizations). Testing, on the other hand, is more scalable and offers error-triggering inputs that facilitate debugging and patching; \tool\ also achieves zero false positives in all of the bugs it found. In line with other works for complex system reliability~\cite{li2019efficient,stoica2023waffle,chen2023push,gu2023acto,sun2022automatic,sun2020testing,gong2023snowcat,gong2025snowplow}, \tool\ adopts testing as its main technique.

\parh{Extension to other JS Engines.}~A potential question is whether \tool\ can generalize to other JS engines beyond the ones evaluated in this work. \tool 's design is not tied to a particular JS engine, and the LLM-guided instrumentation pipeline operates over C++ source and generic ASTs. \tool\ is evaluated on a broad range of JS engines, including the default JS engines from major browsers (V8, SpiderMonkey, JavaScriptCore) and those used in embedded systems and IoT devices (QuickJS, Hermes, Escargot). We believe that extending to other JS engines requires minimal engineering effort.

\parh{LLM Agents for Vulnerability Discovery.}~
LLM agents have begun to
discover vulnerabilities on their own. Big Sleep~\cite{bigsleep} and
Mythos~\cite{mythos} have disclosed bugs in widely deployed software such as
the Linux kernel, FFmpeg, and OpenSSL, and the approach now spans OS
kernels~\cite{knighter2025,li2024enhancing}, web
applications~\cite{stafeev2025yurascanner,ji2025artemis}, smart
contracts~\cite{liu2024propertygpt,sun2024gptscan}, firmware and embedded
systems~\cite{ji2026firmagent}, binaries~\cite{liu2025llm,chen2025clearagent},
reverse engineering~\cite{wong2025decllm,llmlifter,wong2026binrag},
and penetration testing~\cite{deng2024pentestgpt,singer2026incalmo}.
Unlike \tool, which invokes the model once offline to synthesize
instrumentation, these systems keep the LLM in the loop as the reasoner
that proposes and validates candidate vulnerabilities.
To see whether this role difference matters in practice, we asked Codex
to search QuickJS for vulnerabilities and produce runnable proofs of 
concept. The search yielded one genuine bug, which \tool\ had also 
found, alongside several candidates that manual triage ruled out. 
This suggests the two approaches are complementary, and we leave their
combination to future work.

\section{Related Work}
\label{sec:related}

\parh{JavaScript Engine Fuzzing.}~
LangFuzz~\cite{holler2012fuzzing}, Jsfunfuzz~\cite{jsfunfuzz},
CodeAlchemist~\cite{han2019codealchemist}, Montage~\cite{lee2020montage},
Superion~\cite{wang2019superion}, DIE~\cite{park2020fuzzing},
and SoFi~\cite{he2021sofi} mutate ASTs of existing test cases;
TokenFuzz~\cite{Salls2021} and CovRL-Fuzz~\cite{eom2024fuzzing}
mutate at the token level;
TemuJS~\cite{wong2025extraction} performs template-based mutation.
Fuzzilli~\cite{gross2023fuzzilli} defines a custom IR (FuzzIL) that captures JS semantics and performs mutations and synthesis on top of it.
OptFuzz~\cite{wang2024optfuzz} improves fuzzing efficiency through path coverage.
A separate line of work targets logic bugs through better testing oracles like conformance
testing~\cite{ye2021automated,park2021jest}, differential
testing~\cite{wang2023fuzzjita,bernhard2022jit,dumpling}, and JIT
compiler validation~\cite{kwon2024translation}.
\tool\ provides state-sensitive feedback that captures a
dimension that standard coverage metrics may miss,
steering the fuzzer toward more diverse runtime behaviors;
it is thus orthogonal to these works.

\parh{Language Virtual Machine Testing.}~
JVM fuzzers have targeted bytecode validity via coverage-guided
mutation~\cite{chen2016coverage} and rule-based
generation~\cite{chen2019deep}, leveraged historical bug patterns to
guide mutant synthesis~\cite{zhao2022history,zhao2024program},
and focused on JIT-specific bugs through optimization-activating
mutators~\cite{wu2023jitfuzz}, joint program--option
mutation~\cite{jia2023detecting}, and cross-tier differential
testing~\cite{li2023validating}.
JAttack~\cite{zang2023jattack} and LeJit~\cite{zang2024java} use
template-based generation to trigger JVM bugs, but their templates
encode fixed syntactic patterns. In contrast, \tool\ targets the
engine's internal state directly by instrumenting state expressions
identified from developer artifacts, providing semantic
feedback orthogonal to code coverage.
Beyond JVMs, other LVMs like WebAssembly 
runtimes~\cite{waltz,wasit2025,liu2023exploring},
BPF verifiers~\cite{3691938.3691971,10.1145/3658644.3690237,10.1145/3731569.3764797}, and Ethereum VMs~\cite{evmfuzzer2019,ma2023loki,chen2023tyr,yang2021finding}, have also been fuzzed.
It would be promising to extend \tool\ to complement the testing
of other LVMs by providing state-sensitive feedback to explore
more diverse program behavior.

\parh{LLM-Assisted Fuzzing.}~
LLMs have shown strong capabilities in code understanding and
program analysis~\cite{gpt51,iclrinvariants2023,knighter2025}. 
A growing line of work applies LLMs to support 
fuzzing~\cite{kernelgpt2025,ragkernelfuzz,hlpfuzz,
zhang2026llm,park2026agentic,tu2026cottontail}, 
with most approaches using LLMs to
directly synthesize test inputs for the target
program~\cite{whitefox2024,fuzz4all2024,hlpfuzz} or to construct 
struct-aware harnesses or grammar that constrain the input search 
space~\cite{elfuzz2025,harnessgen2025,chatafl}.
\tool\ takes a complementary approach: rather than using LLMs
to generate inputs or harnesses, it leverages LLMs to synthesize instrumentation probes as guidance for the fuzzer.

\section{Conclusion}
\label{sec:conclusion}

We present \tool, a state-aware fuzzing framework that finds JavaScript engine
bugs beyond the reach of edge coverage. By using an LLM agent to mine developer
artifacts and synthesize lightweight instrumentation, \tool\ turns semantically
meaningful internal states into fuzzing feedback without manual, per-engine
annotation.
Across six production JavaScript engines, \tool\ uncovered 68 bugs, 
at least 35 of them with security implications.
These results demonstrate the effectiveness and real-world impact of \tool\
and offer a promising direction for testing and securing other large,
stateful software systems.

\section*{Acknowledgment}
We would like to thank the anonymous reviewers and our shepherd, Andi Quinn, for their valuable feedback. The HKUST authors were supported in part by a grant from the Research Grants Council of the Hong Kong Special Administrative Region, China HKUST C6004-25G and 16214723.

\bibliographystyle{plain}
\bibliography{ref}

\end{document}